# Anisotropic Cloud String Cosmological Model With Bianchi Type-I Space-Time in General Relativity

Kangujam Priyokumar Singh[1], Meher Daimary[2, *]

[1,2]Department of Mathematical Sciences, Bodoland University, Kokrajhar, Kokrajhar-783370, BTC, Assam, India

*Corresponding author: meherdaimary@gmail.com

**Abstract:** Anisotropic cloud string cosmological model with Bianchi type-I space-time is investigated in the context of general relativity. The exact solutions of Einstein's field equations are obtained with when the source for energy momentum tensor is generated by a cloud of strings with particles attached to them. The dynamical and physical properties of the model universe are discussed by comparing with the present observational findings. The interesting feature obtained here is that our model universe starts with a big bang and as time passes both particle density $\rho_p$ and string tension density λ decreases with expansion of our Universe so that in the late time string vanishes and thus leaving only the particles.

## 1. Introduction

The study of the beginning of our Universe and early stage of its formation is still an interesting area of research to discover its unknown phenomenon that have yet to be observed to investigate about evolution of the Universe. So, cosmologists have taken considerable interest to understand of its future evolution and comprehension of the past and present situation of the Universe. But different minds provide different opinions about the Universe, so till today we cannot make a final statement about its origin and evolution with strong evidence. So, further investigation is our moral responsibility to discover the unknown phenomenon of the whole Universe and many other mysterious particles that have yet to be discovered to make a clear statement of our Universe. The origin and early evolution of universe can be explained by string theory. The general relativistic study of string was started by Stachel [1] and Letelier [2]. In recent years, many prominent authors are interested in cosmic strings in the frame-work of general relativity as they are believed to play a key role in early stages evolution of Universe [3, 4] and to density perturbation thereby creating galaxies [5, 6].

The strings are nothing but one dimensional hypothetical topological defect which occurs during phase transition from a temperature below some critical temperature at early evolution of Universe. The presence of strings during early days of Universe does not contradict the present day's observation. These strings posses stress energy and interact with gravitational field. However in the present days, strings are undetectable but their existence leads to anisotropy in space-time. The string can describe simultaneously the nature and fundamental configuration of early Universe. String theory provides us with a single theoretical structure, where all the matters and forces are united and describe the early evolution of Universe in terms of (vibrating) strings rather than particles. Due to the key role of strings in describing the evolution of early stage of our Universe, in recent times, many authors extensively studied the string cosmological models of our Universe. After the big-bang, there is a symmetry breaking during the phase transition of early stages of Universe and these strings arises when the cosmic temperature reaches to a temperature lower than some critical temperatures according to GUT (grand unified theories) [3-5,7-9] . Therefore, in order to study about the early Universe, strings can play a crucial role. Also, the study of gravitational effects arising from the cosmic strings is treated as one of the interesting topics.

Various literatures reveal that on a large scale, FRW (Friedmann-Robertson-Walker) space-time describes only isotropic, homogeneous Universe expanding with accelerated rate. But Bianchi-type string cosmological models include four dimensional spatially homogeneous space-times which satisfy necessary conditions at its lowest-order string beta-function equations. Disposing with the idea of spatial isotropy, Bianchi-type space-times with string generalize all possible FRW models and also agrees with those asymptotic or less than $SO_{(3)}$ isotropy [10, 11]. As such, they are treated as one of the best models available for understanding of anisotropy and its impact on dynamics of the early Universe, well before attainment of isotropy, the present day's observational state. Bianchi type cosmological models represent the homogeneous and anisotropic Universe, where the isotropy process of them may also be studied with the passage of time. Also, in the theoretical perspective, the anisotropic Universes have possessed a greater generality than the isotropic model universes.

Many researchers like [12-17] have studied different string cosmological models. Considering Bianchi type-II, -VI0, -VIII and -IX space-times, [18, 19] have studied the Letelier string cosmological model in different context to obtain the exact solutions of the model. Investigating LRS Bianchi type-I metric, Xing [20] obtained an exact solution string cosmological model with bulk viscosity by assuming the coefficient of bulk viscosity as a power function of the energy density. In the spherical symmetric space-time, Yavuz [21] examined charged strange quark matter attached to the string cloud and clearly shown that one-parameter group of conformal motions has existed. Yilmaz [22] obtained the Kaluza-Klein cosmological solutions for quark matter attached to the string cloud in the context of general relativity. Investigating a Bianchi type-V space-time in a scalar-tensor theory, Rao [23] obtained an exact perfect fluid cosmological model based on Lyra manifold. Tripathi [24] studied an anisotropic and spatially homogeneous Bianchi type-VI0 space-time and obtained string cloud cosmological models in Saez-Ballester Scalar-Tensor theory. Adhav et al. [25] obtained string cosmological models in Brans-Dicke theory of gravitation, where they have solved the field equations by using the condition that the sum of the tension density and energy density is zero. Recently, Pradhan [26] studied a class of anisotropic and homogeneous Bianchi type-V cosmological models with massive strings.

Motivated from the above literatures, in this paper we study string cosmological model with particles attached to them in Bianchi type-I space-time. By taking the average scale factor of the Universe as an integrating function of time as $a = (t^k e^t)^{\frac{1}{l}}$, we attempt to obtain exact solutions of Einstein's field equations to investigate further about the role of strings in the early stage of the Universe. The work, presented here is somewhat different from the earlier findings in many areas that can be seen from our discussions. The paper is organized as follows: In Section 2, the metric and the field equations are presented; Section 3, deals with the solution of field equations; Physical and kinematical properties of our model universe are presented in Section 4; In Section 5 concluding remarks are presented.

## 2. Metric and Field Equations

We consider the spatially homogeneous and anisotropic Bianchi type-I metric in the form

$$ds^2 = -dt^2 + A^2 dx^2 + B^2 dy^2 + C^2 dz^2, \tag{1}$$

where $A, B$ and $C$ are the metric functions of cosmic time $t$ only.

The energy-momentum tensor for a cloud string has the form

$$T_i^j = \rho v_i v^j - \lambda x_i x^j, \tag{2}$$

where $v_i$ and $x_i$ satisfy the conditions

$$v^i v_i = -x^i x_i = -1, \tag{3}$$

$$u^i x_i = 0, \tag{4}$$

where, $\rho$ being the rest energy density for a cloud of strings with particles attached to them. $\rho = \rho_p + \lambda$, $\rho_p$ being the particle density of the configuration and $\lambda$, the tension density of the string. $x_i$ is a unit space-like vector representing the direction of strings so that $x^2 = 0 = x^3 = x^4$ and $x^1 \neq 0$, and $v_i$ is the four velocity vector satisfying the following conditions

$$g_{ij} v^i v^j = -1, \qquad (5)$$

In the present scenario, the co-moving co-ordinates are taken as

$$v^i = (0,0,0,1) \qquad (6)$$

We choose the direction of string parallel to $x$-axis so that

$$x^i = (\tfrac{1}{A},0,0,0) \qquad (7)$$

The Einstein's field equations (with $\frac{8\pi G}{c^4} = 1$)

$$R_i^j - \tfrac{1}{2} R g_i^j = -T_i^j, \qquad (8)$$

where $R_i^j$ is the Ricci tensor, $R = g^{ij} R_{ij}$ is the Ricci scalar.

The Einstein's field equations (8) together with (2) for the line-element (1) lead to the following system of equations:

$$\frac{\ddot{B}}{B} + \frac{\ddot{C}}{C} + \frac{\dot{B}\dot{C}}{BC} = \lambda \qquad (9)$$

$$\frac{\ddot{A}}{A} + \frac{\ddot{C}}{C} + \frac{\dot{A}\dot{C}}{AC} = 0 \qquad (10)$$

$$\frac{\ddot{A}}{A} + \frac{\ddot{B}}{B} + \frac{\dot{A}\dot{B}}{AB} = 0 \qquad (11)$$

$$\frac{\dot{A}\dot{B}}{AB} + \frac{\dot{B}\dot{C}}{BC} + \frac{\dot{A}\dot{C}}{AC} = \rho \tag{12}$$

where an over dot stands for the first and double over dot for the second derivative with respect to cosmic time $t$.

The spatial volume for the model (1) is given by

$$V^3 = a^3 = ABC, \tag{13}$$

We define $V = (ABC)^{\frac{1}{3}}$ as the average scale factor so that the Hubble parameter in anisotropic models may be defined as

$$H = \frac{\dot{V}}{V} = \frac{1}{3}\left(\frac{\dot{A}}{A} + \frac{\dot{B}}{B} + \frac{\dot{C}}{C}\right), \tag{14}$$

We define the generalized mean Hubble's parameter $H$ as

$$H = \frac{1}{3}(H_1 + H_2 + H_3), \tag{15}$$

where $H_1 = \frac{\dot{A}}{A}, H_2 = \frac{\dot{B}}{B}$ and $H_3 = \frac{\dot{C}}{C}$ are the directional Hubble's parameters in the directions of $x, y$ and $z$ respectively.

The deceleration parameter $q$, which is defined as

$$q = -\frac{V\ddot{V}}{\dot{V}^2}, \tag{16}$$

The physical quantities of observational interest in cosmology i.e. the expansion scalar $\theta$, the shear scalar $\sigma^2$ and the average anisotropy parameter $\Delta$ are defined as

$$\theta = v^i_{;i} = 3\frac{\dot{V}}{V} = \left(\frac{\dot{A}}{A} + \frac{\dot{B}}{B} + \frac{\dot{C}}{C}\right), \tag{17}$$

$$\sigma^2 = \frac{1}{2}\left(\sum_{i=1}^{3} H_i^2 - 3H^2\right) \tag{18}$$

$$\Delta = \frac{1}{3}\sum_{i=1}^{3}\left(\frac{\Delta H_i}{H}\right)^2, \tag{19}$$

where $\Delta H_i = H_i - H$ (i = 1, 2, 3)

### 3. Solution of the Field Equations

The field equations (9) − (12) are a system of four equations in five unknown parameters $A, B, C, \rho$ and $\lambda$. One additional constraint relating these parameters is required to obtain explicit solutions of the system. So, based on the works of Thorne [27], and from the observation of the velocity-red-shift relation for extragalactic sources which suggests that the Hubble expansion of the Universe is isotropic to 30% as suggested by Kantowski and Sanchs [28], Kristian and Sanchs [29]. More precisely, the red-shift places the limit $\frac{\sigma}{H} \leq 0.30$, where $\sigma$ is shear and $H$ is Hubble constant. Also, Collins et al. [30] have pointed out that for spatially homogeneous metric, the normal congruence to the homogeneous hyper-surface that satisfies the condition $\frac{\sigma}{\theta}$ is constant, where $\theta$ is the expansion in the model. So we assume the expansion scalar $\theta$ is proportional to the shear scalar $\sigma$ which leads the relation

$$A = B^m, \tag{20}$$

where $m$ is constant.

From (11) and (12), we get

$$B = nC, \tag{21}$$

The average scale factor as an integrating function of time is [31] given by

$$a = (t^k e^t)^{\frac{1}{l}}, \tag{22}$$

where $k$ and $l$ are positive constants.

Using (20), (21) and (22) in equation (13), we get

$$B = n^{\frac{1}{m+2}}(t^k e^t)^{\frac{3}{l(m+2)}} \qquad (23)$$

Using (23) in equation (20), we get

$$A = n^{\frac{m}{m+2}}(t^k e^t)^{\frac{3m}{l(m+2)}} \qquad (24)$$

Using (23) in equation (21), we get

$$C = n^{-\frac{m+1}{m+2}}(t^k e^t)^{\frac{3}{l(m+2)}}, \qquad (25)$$

Using equations (23), (24) and (25), the metric equation (1) takes the form

$$ds^2 = -dt^2 + n^{\frac{2m}{m+2}}(t^k e^t)^{\frac{6m}{l(m+2)}}dx^2 + n^{\frac{2}{m+2}}(t^k e^t)^{\frac{6}{l(m+2)}}dy^2 + n^{-2\left(\frac{m+1}{m+2}\right)}(t^k e^t)^{\frac{6}{l(m+2)}}dz^2 \qquad (26)$$

Equation (26) represents Bianchi type-I cosmological model in general relativity with time dependent deceleration parameter.

### 4. Physical and Kinematical Properties of the Model

For the cosmological model (26), the physical quantities such as proper volume $V$, Hubble parameter $H$, expansion scalar $\theta$, mean anisotropy parameter $\Delta$, shear scalar $\sigma^2$, energy density $\rho$, particles density $\rho_p$, tension density of the string $\lambda$ and deceleration parameter $q$ are obtained as follows:

The proper volume $V$ is given by

$$V = (t^k e^t)^{\frac{3}{l}}, \qquad (27)$$

The Hubble parameter $H$ is in the form

$$H = \frac{1}{l}(1 + kt^{-1}), \qquad (28)$$

The expansion scalar $\theta$ is

$$\theta = \frac{3}{l}(1 + kt^{-1}), \tag{29}$$

The mean anisotropy parameter $\Delta$ is given by

$$\Delta = 2\left(\frac{m-1}{m+2}\right)^2 = Constant(\neq 0 for m \neq 1), \tag{30}$$

The shear scalar $\sigma^2$ is

$$\sigma^2 = \frac{3(m-1)^2}{l^2(m+2)^2}(1 + kt^{-1})^2, \tag{31}$$

From (29) and (31), we have

$$\lim_{t \to \infty} \frac{\sigma^2}{\theta^2} = \frac{(m-1)^2}{3(m+2)^2} = Constant(\neq 0 for m \neq 1), \tag{32}$$

The energy density $\rho$ is given by

$$\rho = \frac{9(2m+1)}{l^2(m+2)^2}(1 + kt^{-1})^2, \tag{33}$$

The particle density $\rho_p$ attached to the string is

$$\rho_p = \frac{3}{l^2(m+2)}[6(1 + kt^{-1})^2(m - 1) + kl(m + 2)t^{-2}], \tag{34}$$

The tension density $\lambda$ of the string is obtained by

$$\lambda = \frac{3}{l^2(m+2)^2}[9(1 + kt^{-1})^2 - kl(m + 2)t^{-2}], \tag{35}$$

The deceleration parameter $q$ is

$$q = \frac{lk}{(t+k)^2} - 1, \tag{36}$$

Taking $k = 1, l = m = 3, l = m = 2, l = m = 1$ the variation of some of the parameters are shown below:

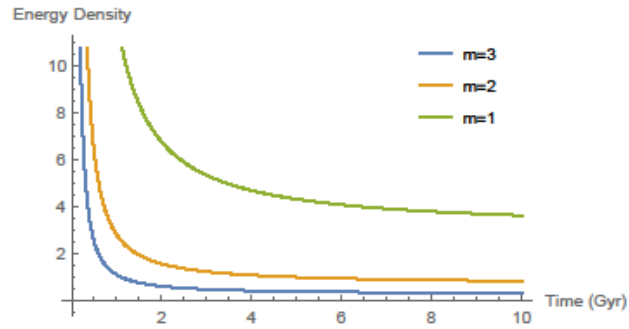

Figure 1: Variation of Energy Density ρ vs. Time t in Gyr.

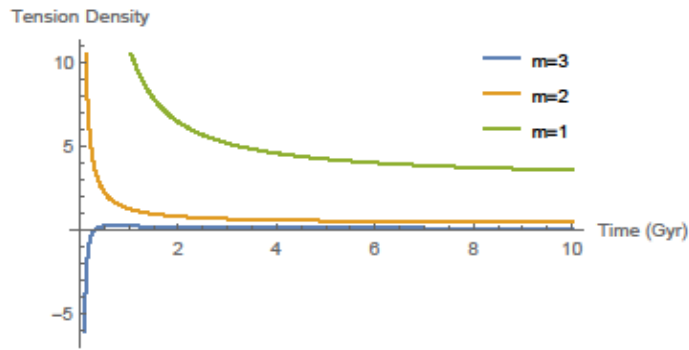

Figure 2: Variation of Tension Density λ vs. Time t in Gyr.

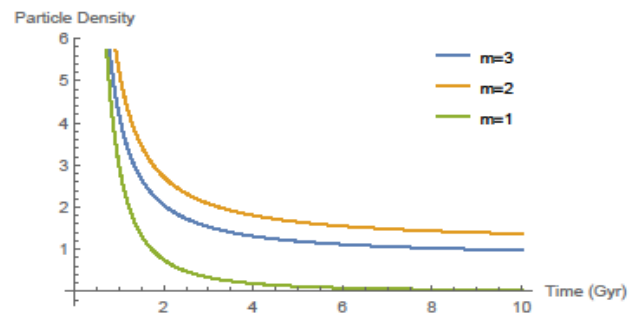

Figure 3: Variation of Particle Density $\rho_p$ vs. Time t in Gyr.

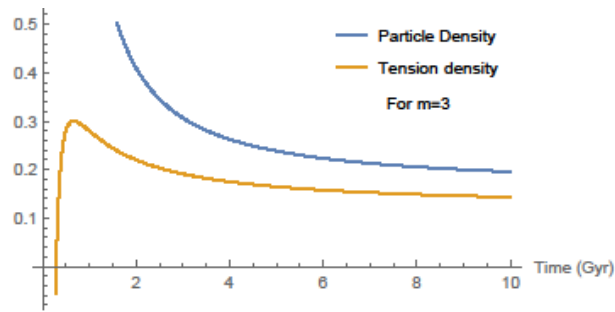

Figure 4: Variation of Both Particle Density $\rho_p$ and Tension Density λ vs. Time t in Gyr.

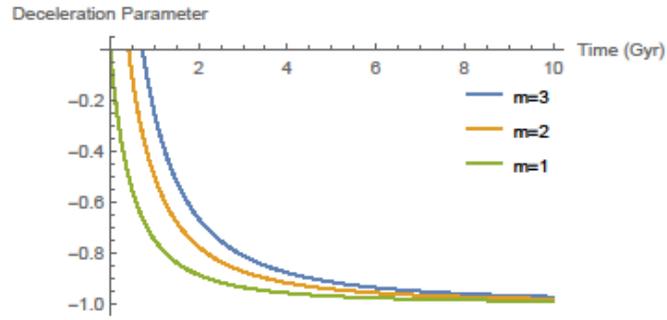

Figure 5: Variation of Deceleration Parameter q vs. Time t in Gyr.

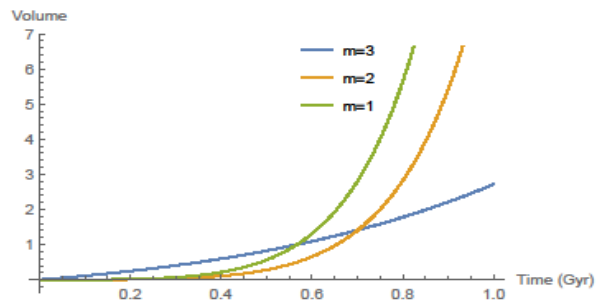

Figure 6: Variation of Volume V vs. Time t in Gyr.

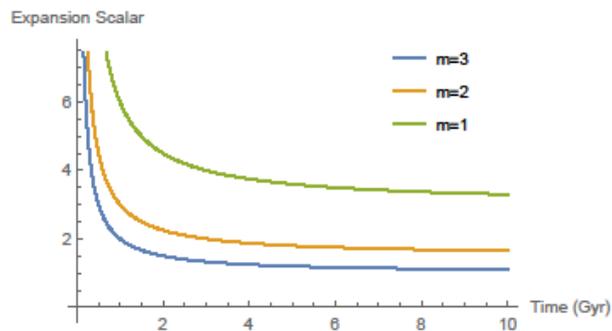

Figure 7: Variation of Expansion Scalar θ vs. Time t in Gyr.

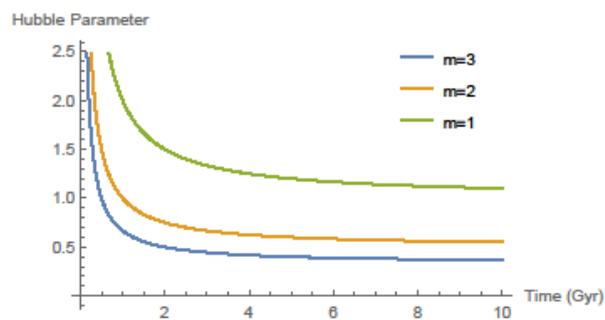

Figure 8: Variation of Hubble Parameter H vs. Time t in Gyr.

The expression for energy density ρ given by equation (33) shows that the energy density $\rho$ is a decreasing function of cosmic time $t$ and satisfy the energy condition $\rho \geq 0$ for all $m \geq -\frac{1}{2}$. The variation of energy density against cosmic time is presented in Figure-1, which also shows that it is a decreasing function of time t and initially when $t \to 0$ then $\rho \to \infty$, so it has an initial singularity.

The equation (35) shows that the string tension density is positive, $\lambda \geq 0$ for all values of cosmic time t but initially when $t \to 0$, $\lambda$ is very large and just after that it becomes a decreasing function of cosmic time $t$ and finally tends to a very small positive quantity. The variation of string tension density $\lambda$ against cosmic time $t$ is presented in Figure-2.

The variation of particle density $\rho_p$ against cosmic time t is shown in Figure-3. From the equation (34) and from the nature of the graph, we see that the particle density $\rho_p$ is a decreasing function of cosmic time t. Initially when $t \to 0$, the value of particle density $\rho_p$ is very large and it has an initial singularity when $t = 0$. The value of $\rho_p$ decreases with increase of time and approaches to a constant value at infinite time which shows that there will remain a finite number of particles in our Universe. This may corresponds to the matter dominated era and decoupling era of radiation. The initial singularity of our Universe is of the point type.

Also, it can be seen that initially when $t \to 0$, both the particle density $\rho_p$ and the string tension density $\lambda$ tend to infinity which indicates that our Universe starts with big bang and as the time progresses both $\rho_p$ and $\lambda$ decreases with the expansion of our universe. From the comparative variation of particle density $\rho_p$ and string tension density $\lambda$ against cosmic time as presented in Figure-4, it can be seen that $|\lambda| < |\rho_p|$ and string tension density $\lambda$ vanishes more rapidly than the particle density $\rho_p$ which describes that in the late time string vanishes, leaving particles only.

The expression (36) shows that the deceleration parameter $q$ is a decreasing function of time $t$. Initially at $t = 0$, the deceleration parameter $q$ is positive which changes sign from positive to negative with the increase of time and at infinite time it tend to -1 which indicates that the proposed model Universe has a transition from decelerating to accelerating phase. It can be confirmed from the graph of deceleration parameter $q$ versus cosmic time $t$ which is presented in Figure-5. It means that this model is found to be expanding with time. Thus our model represents an interesting model Universe to be studied for a desirable feature of a meaningful string model.

From the expression (27) of spatial volume V, we observed that it is an increasing function of cosmic time $t$ increasing exponentially and evolves with zero volume at $t = 0$ and it becomes infinite as $t \to \infty$. Figure-6 depicts this behaviour of volume V. Also, when $t = 0$, the scale factors A(t), B(t) and C(t) are found to be infinite.

From equations (29) and (28) and their respective graphical presentations shown in Figure-7 & Figure-8 depict that at the initial epoch of cosmic time $t = 0$, both the expansion scalar $\theta$ and the Hubble's parameter $H$ are infinite and are decreasing functions of time t which are approaching to finite values at $t \to \infty$. The initial singularity of the model is of the point type which supports the findings of Ref. [32]. Since $\frac{\sigma^2}{\theta^2}$ = constant for $m \neq 1$, so the model does not approach isotropy. Also, from the expression of Hubble's expansion factor (28), we found that $\frac{dH}{dt}$ is negative which indicates that our model corresponds to an expanding universe which starts evolving at t = 0 and is expanding with an accelerated rate.

## 5. Concluding Remarks

In this paper, we attempt to explain the behaviour of a cloud string to describe some of unknown phenomenon of Universe. Bianchi type-I string cosmological model with anisotropic background space-time structure is studied with the use of certain physical assumptions which are matching with the present observational findings in the context of general theory of relativity. The model represents to have anisotropic phase throughout the evolution of the universe which is in agreement with the present observational data made by COBE (Cosmic Background Explorer) and WMAP (The Wilkinson Microwave Anisotropy Probe) where they found small anisotropy in the microwave background radiation. Also, the model represents an exponentially expanding Universe which starts with big-bang at cosmic time $t = 0$ with zero volume and expands with acceleration. Our model satisfies the energy conditions $\rho \geq 0$ and $\rho_p \geq 0$. The particle density and string tension density are comparable whereas the string tension density vanishes more rapidly than the particle density, so in the late time our model represents a matter dominated Universe, that the present day's observational data.

**Acknowledgement**

The authors are thankful to the Council of Scientific and Industrial Research (CSIR), India for funding under the sanction Order No. **25(0279)/18/EMR-II** of dated **4th April, 2018**, to carry out this work successfully. The authors are also thankful to the honourable referees for their valuable comments and suggestions.